# On the Self-Consistent Event Biasing Schemes for Monte Carlo Simulations of Nanoscale MOSFETs

**Sharnali Islam[1], Mihail Nedjalkov[2], and Shaikh Ahmed[1]**

[1]Department of Electrical and Computer Engineering
Southern Illinois University at Carbondale
1230 Lincoln Drive, Carbondale, IL 62901, USA.
Phone: (618) 453-7630, Fax: (618) 453-7972, E-mail: ahmed@siu.edu

[2] Institute for Microelectronics, Technical University of Wien, Gußhausstraße, Austria
Phone: +43 1 58801-36048, E-mail: nedialkov@iue.tuwien.ac.at

*Abstract*—

Different techniques of event biasing have been implemented in the particle-based Monte Carlo simulations of a 15nm *n*-channel MOSFET. The primary goal is to achieve enhancement in the channel statistics and faster convergence in the calculation of terminal current. Enhancement algorithms are especially useful when the device behavior is governed by rare events in the carrier transport process. After presenting a brief overview on the Monte Carlo technique for solving the Boltzmann transport equation, the basic steps of deriving the approach in presence of both the initial and the boundary conditions have been discussed. In the derivation, the linearity of the transport problem has been utilized first, where Coulomb forces between the carriers are initially neglected. The generalization of the approach for *Hartree* carriers has been established in the iterative procedure of coupling with the Poisson equation. It is shown that the weight of the particles, as obtained by biasing of the Boltzmann equation, survives between the successive steps of solving the Poisson equation.

*Index*—event biasing, Monte Carlo simulations, nanoscale MOSFET, statistical enhancement, Boltzmann transport equation.

# 1    Introduction—Semiclassical Electron Transport

The semiclassical electron transport in semiconductor materials and devices is governed by the Boltzmann transport equation, which expresses the global non-equilibrium distribution function, $f(\mathbf{r},\mathbf{k},t)$, in terms of the local equilibrium distributions under various applied and built-in forces. In its most general form the BTE reads:

$$\frac{\partial f}{\partial t} + \mathbf{v} \cdot \nabla_r f + \frac{\mathbf{F}}{\hbar} \cdot \nabla_{\mathbf{k}} f = \frac{\partial f}{\partial t}\bigg|_{scatt}, \qquad (1)$$

where $v$ is the carrier group velocity. The terms on the left-hand side represent the change in the distribution function with respect to time, spatial gradients, and applied fields. The right-hand side represents the dissipation terms in the system, which accounts for the change of the distribution function due to various scattering mechanisms that balance the driving terms on the left. The Boltzmann equation is valid under assumptions of semiclassical transport [1]: (1) *effective mass* approximation taking into account the static quantum effects due to periodicity of the semiconductor crystal. (2) *Born approximation* for the collisions in the limit of small perturbation for the electron-phonon interaction and instantaneous collisions; scattering probability is independent of external forces; no memory effects, i.e. no dependence on initial condition terms; particle interactions are uncorrelated and forces are constant over distances comparable to the electron wave function. (3) The *phonons* are usually treated as being in equilibrium, although the condition of non-equilibrium phonons may be included through an additional phonon transport equation. Boltzmann transport equation finds its application in different fields in science and engineering such as nuclear reactor design, radiation shielding calculations, radiative transfer in stellar atmospheres, semiconductor device design, radiation oncology, and high energy physics [2]. Analytical solutions of the Boltzmann equation are possible only under very restrictive assumptions. There are mainly two classes of computational techniques that are used to solve the transport equation—(1) In the first class, in *numerical/deterministic methods*, the transport equation is discretized using a variety of methods and then solved directly or iteratively. Direct numerical methods are limited by the complexity of the equation, which in the complete 3D time-dependent form requires seven independent variables for time, space and momentum. Different types of discretization schemes give rise to different deterministic methods [3] [4], such as discrete ordinates ($S_N$), spherical harmonics ($P_N$), collision probabilities, nodal methods, and others. (2) The second class of techniques, named *Monte Carlo* methods, constructs a *stochastic model* in which the expected value of a certain random variable is equivalent to the value of a physical quantity to be determined [5] [6] [7] [8]. The expected value is estimated by the average of many independent samples representing the random variable. Random numbers, following the distributions of the variable to be estimated, are used to construct these independent samples. There are two different ways to construct a stochastic model for Monte Carlo calculations. In the first case the physical process is stochastic and the Monte Carlo calculation involves a computational simulation of the real physical process. This is achieved by tracing the trajectories of individual carriers as they are accelerated by the electric field and experience random scattering events. The particle movements between scattering events are described by the laws of classical mechanics, while the probabilities of the various scattering processes and the associated transition rates are derived from quantum mechanical calculations. The randomness of the events is treated in terms of computer generated random numbers, distributed in such a way as to reflect these probabilities. In the other case, a stochastic model is constructed artificially, such as the solution of deterministic equations by Monte Carlo [9]. Both the deterministic and the Monte Carlo stochastic methods have computational errors. Deterministic methods are computationally fast with less accuracy; whereas Monte Carlo methods are computationally slow yet arbitrarily accurate. *A great advantage of Monte Carlo methods is that it provides a unique insight into the underlying device physics*. To-date, most semiconductor applications have been based on stochastic solution methods (particle-based Monte Carlo), which involve the simulation of particle trajectories rather than the direct solution of partial differential equations.

## 2     The Monte Carlo Method

The Monte Carlo (MC) method was originally used and devised by Fermi, Von Neumann and Stanislaw Ulam to solve the BTE for transport of neutrons in the fissile material of the atomic bomb during the Manhattan Project of World War II [10]. Since these pioneering times in the mid 1940's, the popularity and use of the MC method has grown with the increasing availability of faster and cheaper digital computers. Its application to the specific problems of high-field electron transport in semiconductors is first due to Kurosawa [11] in 1966.

Shortly afterwards the Malvern, UK, group [12] provided the first wide application of the method to the problem of the *Gunn effect* in GaAs. Applications to Si and Ge boomed in the 1970s, with an extensive work performed at the University of Modena, Italy. In the mid-1970s, a physical model of silicon was developed, capable of explaining major macroscopic transport characteristics. The used band structure models were represented by simple analytic expressions accounting for nonparabolicity and anisotropy. The review articles by C. Jacoboni and L. Reggiani [13], and by Peter J. Price [14] provided a comprehensive and deeper historical and technical perspective.

The Monte Carlo method is the most popular method used to solve the Boltzmann transport equation without any approximation to the distribution function. In the Monte Carlo method, particles are used to represent electrons (holes) within the device. For *bulk* simulations, the momentum and energy of the particles (electrons and/or holes) are continuously updated. For *device* simulations, the real space position of the particle is updated as well. As time evolves, the updated momentum (and corresponding energy) is calculated from the various forces applied on the particle for that time step. The general concept of a Monte Carlo simulation is that the electrons (holes) are accelerated by an electric field until they reach a predetermined scattering time, defines as $t_{\text{scat}}$. At the scattering time, a scattering mechanism is randomly chosen based on its relative frequency of occurrence. The basic steps in a typical Monte-Carlo particle-based *device* simulation scheme include [see Refa. [15] [16] [17]]:

1. Initialization.
2. Field (Poisson) equation: determine forces on electrons.
3. Electron dynamics (free-flight and scattering):
    (a) Accelerate the electrons.
    (b) Determine whether the electron experiences a scattering/collision.
    (c) If electron scatters, select the scattering mechanism.
    (d) Update the electron position, energy and wavevector.
4. Charge assignment.
5. Compute measurable quantities such as average energy or average velocity of the ensemble.
6. Repeat steps 2-5 for each iteration.

The *initialization* includes the calculation of various material parameters such as the electronic band structure, scattering rates and (particularly for *device* simulation) the definition of the computational domain elements and initial electron distributions. In order to determine device characteristics, such as drain current, it is necessary to solve the carrier transport equations using the device parameters (including doping concentration, channel length, channel width, and oxide thickness) and boundary conditions (such as terminal voltages) for the particular device being simulated. The device is divided into grids/cells by a numerical discretization technique (usually finite difference method). For each dimension, maximum number of grid points is set as a program parameter so that array sizes can be properly allocated. Provision may be kept for nonuniform mesh spacing along the different dimensions of the device. The source and drain contact charges are then calculated typically using the doping density and the mesh size. Then, the Poisson equation is solved for the applied gate bias only (keeping source/drain/substrate bias equal to zero) and the resulting potential distribution is used to populate/initialize carriers in each of the cells in the active region. It is always convenient to begin the simulation in thermodynamic equilibrium where the solution is known. When an electron is added to the device active region, the real space components of the electron are determined based upon the position of the node where it is being added. Under charge-neutral conditions, the total number of free carriers within the device must equal the total number of ionized dopants. In the non-equilibrium step, the remaining boundary biases are applied at different contacts/terminals (source/drain/substrate). The program control is thus transferred to the Monte Carlo iterative transport/dynamics kernel. Here, for each pre-selected time step (typically fractions of femtosecond) the carriers undergo the free-flight-scatter sequence. During a time interval, $\Delta t$, each electron is accelerated according to Newton's second law of motion. For a semiconductor with *ellipsoidal* constant energy surfaces, such as silicon, the effective mass is a tensor quantity. Using Hering-Vogt transformation [15]

$$k_i^* = \sqrt{\frac{m^*}{m_i}} k_i \quad i = x, y, z \tag{2}$$

where $m_i$ is the effective mass of the particle in the $i^{th}$ ($x$, $y$, or $z$) direction, one can make a transformation from $k$-space to $k^*$-space in which the constant energy surfaces are *spherical*. In the original $k$-space, due to the mass anisotropy, the constant energy surfaces are ellipsoidal. In order to determine the time between scattering events, the electron momentum at the time of scattering must be known, since the scattering rate is a function of the electron momentum. The scattering rates for the various scattering mechanisms included into the model are then tabulated (in collisions per second) as a function of energy. Once a scattering mechanism is chosen, a new state of the electron must be determined by the type of scattering event selected. For elastic scattering, the electron energy is unchanged, and a new *k*-vector is randomly chosen. For inelastic scattering processes, such as optical phonon scattering, the electron energy must be increased or decreased by the phonon energy depending on whether absorption or emission process occurs. The final **k**-vector is then randomly chosen. After a new *k*-vector is chosen, a new scattering time is determined. The electron is then accelerated for the remainder of the time interval, or until it scatters. All of the electrons are accelerated and scattered until they reach the interval time $\Delta t$. Also, to simulate the *device*, the boundaries must be treated properly. The Ohmic contacts are often assumed to be perfect absorbers, so carriers that reach them simply exit the device. At the end of each timestep, thermal electrons are injected from the contacts to maintain space-charge neutrality therein. The noncontacted free surfaces are treated as reflecting boundaries. For field-effect transistors, roughness at the surface of the channel can cause scattering. A simple approach is to treat some fraction of the encounters with the surface as specular scattering events and the remainder as diffusive scattering events. The specific fraction is usually selected to match transport measurements, such as the low-field mobility as a function of the transverse electric field. A carrier deletion scheme is also implemented at this stage. When completed, charges are assigned to the nearest node points using the charge-assignment method. The charges obtained from the EMC simulation are usually distributed within the continuous mesh cell instead of on the discrete grid points. The particle mesh method (PM) is used to perform the switch between the continuum in a cell and discrete grid points at the corners of the cell. The charge assignment to each mesh-point depends on the particular scheme that is used. A proper scheme must ensure proper coupling between the charged particles and the Coulomb forces acting on the particles. Therefore, the charge assignment scheme must maintain zero self-force and a good spatial accuracy of the forces. Poisson equation is then solved to determine the resulting potential distribution within the device. It is important to note that the electron concentration (in an *n*-MOSFET simulation) is *not* updated based upon the potential at the node point. In contrary, the hole concentration at each node point is calculated using the updated node potential with the assumption that the hole quasi-Fermi level is equal to that of the electrons. The error introduced with the assumption that the quasi-Fermi levels for electrons and holes are equal is small since the hole concentration is negligible in the active region of the (*n*-MOSFET) device. The force on each electron is then interpolated from the nearest node points. The resultant electric field is than used to drive the carriers during the free-flight in the next time step. The whole process is repeated for several thousand iterations until the steady state is achieved. At any time during the simulation, key device measurements such as the average carrier density, velocity, and energy versus position are computed by averaging over the particles within each slab of the active region. Lundstrom [16], Tomizawa [17], and Kunikiyo *et al*. [18] discuss the application of this approach to the simulation of two-dimensional transistors. A detailed description of the 3D Monte Carlo device simulator can be found in Ref. [19].

## 3    Current Calculation

The *device output current* can be determined using two different yet consistent methods. *First*, by keeping track of the charges entering and exiting each terminal/contact, the net number of charges over a period of the simulation can be used to calculate the terminal current. The net charge crossing a terminal boundary is determined by [15]

$$Q(t) = e\left(n_{abs}(t) - n_{injec}(t)\right) + \varepsilon \int E_y(x,t) dy, \tag{3}$$

where $n_{abs}$ is the number of particles that are absorbed by the contact (exit), $n_{injec}$ is the number of particles that have been injected at the contact, $E_y$ is the vertical field at the contact. The second term in (3) on the right-hand-side is used to account for the displacement current due to the changing field at the contact. Eq. (3) assumes the contact is at the top of the device and that the fields in the *x* and *z* direction are negligible. The *e* in (3) should be multiplied by the particle charge if it is not unity. The slope of $Q(t)$ versus time gives a measure of the terminal current. In steady state, the current can be found by

$$I = \frac{dQ(t)}{dt} = \frac{e(n_{net})}{\Delta t}, \tag{4}$$

where $n_{net}$ is the net number of particles exiting the contact over a fixed period of time $\Delta t$. *The method is quite noisy, due to the discrete nature of the electrons*. In a *second* method, the sum of the electron velocities in a portion of the channel region of the device is used to calculate the current. The electron current density through a cross-section of the device is given by

$$J = env_d,  \qquad (5)$$

where $v_d$ is the average electron drift velocity and $n$ is the carrier concentration. If there are a total of $N$ particles in a differential volume, $dV = dL \cdot dA$, the current found by integrating (5) over the cross-sectional area, $dA$, is

$$I = \frac{eNv_d}{dL},  \qquad (6)$$

or,
$$I = \frac{e}{dL}\sum_{i=1}^{N} v_x(i),  \qquad (7)$$

where $v_x(i)$ is the velocity along the channel of the $i^{th}$ electron. The device is divided into several sections along the $x$-axis, and the number of electrons and their corresponding velocity is added for each section after each free-flight. The total $x$-velocity in each section is then averaged over several timesteps to determine the current for that section. Total device current can be determined from the average of several sections, which gives a much smoother result compared to counting terminal charges. By breaking the device into sections, individual section currents can be compared to verify that there is conservation of particles (constant current) throughout the device. In addition, sections near the source and drain regions may have a high $y$-component in their velocity and should be excluded from the current calculations. Finally, by using several sections in the channel, the average energy and velocity of electrons along the channel can be observed to ensure proper physical characteristics.

In the three-dimensional particle based Monte Carlo device simulator used in this work, intravalley scattering is limited to acoustic phonons. For the intervalley scattering, both *g*- and *f*-phonon processes have been included. It is important to note that, by group symmetry considerations, the zeroth-order low-energy *f*- and *g*-phonon processes are forbidden. Nevertheless, three zeroth-order *f*-phonons and three zeroth-order *g*-phonons with various energies are usually assumed [12]. This selection rule has been taken into account and two high-energy *f*- and *g*-phonons and two low-energy *f*- and *g*-phonons have been considered. The high-energy phonon scattering processes are included via the usual zeroth-order interaction term, and the two low-energy phonons are treated via a first-order process [20]. The first-order process is not really important for low-energy electrons but gives a significant contribution for high-energy electrons. The low-energy phonons are important in achieving a smooth velocity saturation curve, especially at low temperatures. The phonon energies and coupling constants in this model are determined so that the experimental temperature-dependent mobility and velocity-field characteristics are consistently recovered [21]. At present, impact ionization and surface-roughness scattering are not included in the model. Impact ionization is neglected, as, for the drain biases used in the simulation, electron energy is typically insufficient to create electron-hole pairs. Also, since it is not quite clear how surface roughness scattering can be modeled when carriers are displaced from the interface due to the quantum confinement effects, it is believed that its inclusion is most likely to obscure the quantum confinement effects. Also, band-to-band tunneling and generation and recombination mechanisms have not been included in the simulations.

## 4    Statistical Enhancement: The Self-Consistent Event Biasing Scheme

Statistical enhancement in Monte Carlo simulations aims at reduction of the time necessary for computation of the desired device characteristics. Enhancement algorithms are especially useful when the device behavior is governed by rare events in the transport process. Such events are inherent for sub-threshold regime of device operation, simulations of effects due to discrete dopant distribution as well as tunneling phenomena. Virtually all Monte Carlo device simulators with statistical enhancement use *population control* techniques [22]. They are based on the heuristic idea for splitting of the particles entering a given phase space region $\Omega$ of interest. The alternative idea—to enrich the statistics in $\Omega$ by biasing the probabilities associated with the transport of classical carriers—gives rise to the *event-biasing* approach. The approach, first proposed for the Ensemble Monte Carlo technique (time-dependent problem) [23], has been recently extended for the Single Particle Monte Carlo technique (stationary problem) [24]. In the next section, the basic steps of derivation of the approach in presence of both initial and boundary conditions has been discussed. Utilized is the linearity of the transport problem, where Coulomb forces between the carriers are initially neglected. The generalization of the approach for *Hartree carriers* has been established in the iterative procedure of coupling with the Poisson equation. Self-consistent simulation results are presented and discussed in the last section.

### 4.1 Event biasing

The Ensemble Monte Carlo (EMC) technique is designed to evaluate averaged values $\langle A \rangle$ of generic physical quantities $a$ such as carrier density and velocity given by

$$\langle A \rangle(\tau) = \int dQ A(Q) f(Q) = \int dQ f_0(Q) g(Q). \tag{8}$$

Here $Q = (\mathbf{k}, \mathbf{r}, t)$ and (8) denotes the integration over the phase space and time $t \in (0, \infty)$, and $A = a\theta_\Omega \delta(t - \tau)$ introduces the indicator $\theta_\Omega$ of the phase space domain, where the mean value is evaluated at time $\tau$. Equation (8) is the usual expression for a statistical mean value, augmented by a time integral with the purpose to conveniently approach the formal theory of integral equations. It has been shown that the Boltzmann equation can be formulated as a Fredholm integral equation of a second kind with a free term $f_0$. The latter is determined by the initial condition in evolution problems [22] [25] or, in the case of stationary transport, by the boundary conditions [25]. The second equality in (8) follows from the relationship between an integral equation and its adjoint equation. It shows that the mean value $\langle A \rangle$ is determined by $f_0$ and by the solution of the adjoint Boltzmann equation

$$g(Q') = \int dQ K(Q', Q) g(Q) + A(Q') \tag{9}$$

$$K = S(\mathbf{k}, \mathbf{k}', \mathbf{r}) e^{-\int_{t'}^{t} \lambda(\mathbf{K}(y), \mathbf{R}(y)) dy} \theta_D(\mathbf{r}) \delta(\mathbf{r}' - \mathbf{r}) \theta(t - t') \tag{10}$$

where $S$ is the usual scattering rate from lattice imperfections, $\lambda$ is the total out-scattering rate, $\theta_D$ is the device domain indicator, which is discussed later, $\theta$ is the Heaviside function and the trajectories, initialized by $(\mathbf{k}, \mathbf{r}, t')$, are formulated with the help of the electrical force $\mathbf{F}$ and the velocity $\mathbf{v}$ as

$$\mathbf{K}(t) = k + \int_{t'}^{t} \mathbf{F}(\mathbf{R}(y)) dy, \tag{11}$$

and

$$\mathbf{R}(t) = \mathbf{r} + \int_{t'}^{t} \mathbf{v}(\mathbf{K}(y)) dy. \tag{12}$$

If both, initial $f_i$ and boundary $f_b$ conditions are taken into account, it can be shown that $f_0$ becomes

$$f_0(Q) = f_i(\mathbf{k},\mathbf{r}) e^{-\int_0^t \lambda(\mathbf{K}(y),\mathbf{R}(y))dy} + \int_0^t \mathbf{v}_\perp(\mathbf{k}) f_b(\mathbf{k},\mathbf{r},t_b) e^{-\int_{t_b}^t \lambda(\mathbf{K}(y),\mathbf{R}(y))dy} dt_b . \tag{13}$$

While $f_i$ is defined only at the initial time $t = 0$, the function $f_b$ is defined only at the device boundary $\Gamma$ and for values of $\mathbf{k}$ such that the corresponding velocity inwards the domain, $D$. $\mathbf{v}_\perp$ is the velocity component normal to $\Gamma$ so that a velocity-weighted distribution drives the particle flux, injected into the device at times $t_b \leq t$. $f_0$ in (13) governs both the transient and the stationary behavior of a device. The latter is established in the long time limit, provided that $f_b$ is time independent. Usually $f_b$ is assumed to be the equilibrium distribution function.

A recursive replacement of equation (9) into itself gives rise to the von-Neumann expansion, where the solution $g$ is presented as a sum of the consecutive iterations of the kernel on $A$. If replaced in (8), the expansion gives rise to the following series for $\langle A \rangle$.

$$\langle A \rangle(\tau) = \sum_i \langle A \rangle_i(\tau) \tag{14}$$

Consider the second term in (14) augmented with the help of two probabilities $P_0$ and $P$ to become expectation [24]

$$\langle A \rangle_2 = \int dQ' dQ_1 dQ_2 P_0(Q') P(Q',Q_1) P(Q_1,Q_2) \frac{f_0(Q')}{P_0(Q')} \frac{K(Q',Q_1)}{P(Q',Q_1)} \frac{K(Q_1,Q_2)}{P(Q_1,Q_2)} A(Q_2), \tag{15}$$

value of a random variable (r.v.). It takes values determined by the second row with a probability given by the product in the first row. $\langle A \rangle_2$ is evaluated according to the numerical Monte Carlo theory as follows. $P_0$ and $P$ are used to construct numerical trajectories: (i) $P_0(Q')$ selects the initial point $Q'$ of the trajectory. (ii) $P(Q',Q)$ selects the next trajectory point $Q$ provided that $Q'$ is given. The fraction $W_2$ in front of $A$, called *weight*, is a product of weight factors $\frac{f_0}{P_0}$, and $\frac{K}{P}$ evaluated at the corresponding points $Q_0 \to Q_1 \to Q_2$, selected by application of $P_0 \to P \to P$. The sample mean of $N$ realizations of the r.v., calculated over $N$ trajectories $(Q' \to Q_1 \to Q_2)_n$, $n = 1...N$, estimates the mean value $\langle A \rangle_2$:

$$\langle A \rangle_2 = \frac{1}{N} \sum_{n=1}^{N} (W_2 A)_n$$
$$\langle A \rangle = \frac{1}{N} \sum_{n=1}^{N} (WA)_n \tag{16}$$

The iterative character of the multiple integral (15) has been used to introduce a consecutive procedure for construction of the trajectories. It can be shown that a single trajectory, obtained by successive applications of $P$, contributes to the estimators of all terms in (14) simultaneously i.e. the procedure is generalized in (16) for a direct evaluation of $\langle A \rangle$. Next, one establishes the link between (16) and the EMC technique, which is due to particular choice of the initial, $P_0^B$, and transition, $P^B$, densities. $P^B$, which can be deduced from (10), is a product of the conditional probabilities for free-flight and scattering, associated with the evolution of the real carriers. The ratio $K/P^B$ is then the domain indicator $\theta_D$ which takes values 1 (one) if the trajectory belongs to $D$ and 0 (zero) otherwise. The choice of $P_0^B$ is complicated by the presence of both initial and boundary terms in (13). They decompose (16) into two terms which are evaluated separately as

$$\langle A \rangle = \frac{1}{N_1} \sum_{n=1}^{N_1} (WA)_n + \frac{1}{N_2} \sum_{n=1}^{N_2} (WA)_n. \tag{17}$$

The initial probability $P_0^B$ for each estimator is obtained from $f_i$ and $\mathbf{v}_\perp f_b$ respectively, with the help of two normalization factors: the number of initial carriers $N_i$ and the total number $N_J$ of the injected particles into the device. The ratio $f_0/P_0^B$ for each of the estimators becomes $N_i$ and $N_J$ respectively, and can be eliminated by the choice $N_1 = N_i$ and $N_2 = N_J$. The two sums can be merged back to give

$$\langle A \rangle = \sum_{n=1}^{N_i+N_J} (WA)_n = \sum_{n=1}^{N_\tau} \theta_\Omega(n) a_n. \tag{18}$$

Equation (18) accounts that only trajectories which belong to $D$ give contributions. As only the endpoint of such trajectories matters for the estimator, we speak about particles inside the device. $N_\tau$ is the number of such particles at time $\tau$, and $\theta_\Omega(n)$ is 1 or 0 if the $n$-th particle is inside or outside $\Omega$. All particles have weight unity and evolve as real Boltzmann carriers and the EMC technique for transport problems posed by initial and boundary conditions is recovered. A choice of alternative probabilities is called *event biasing*. Biased can be the probabilities for initial and/or boundary distributions, free-flight duration, type of scattering and the selection of the after-scattering state direction. It can be shown that (18) is generalized to $\langle A \rangle = \sum_{n=1}^{N_\tau^b} W_n \theta_\Omega(n) a_n$ where the position of the $N_\tau^b$ biased particles is accounted in $\theta_\Omega$.

The Boltzmann equation for Coulomb carriers becomes nonlinear via the interaction component $\mathbf{F}(f)(\mathbf{r},t)$ of the electric force. As the results of the previous section are based on the linearity of the integral equations involved, it is no more possible to apply the steps used to derive the event biasing. The solution is sought in the iterative procedure of coupling of the EMC technique with the Poisson equation. The latter is discretized, as stated earlier, by a decomposition of the device region into mesh cells, $\psi_l$. The particle system is evolved in time intervals $\Delta t \square 0.1$ fs. At the end of each time step, at say time $\tau$, the charge density $eC(\mathbf{r}_l,\tau)$ is calculated and assigned to the corresponding grid points. One uses the relation between $C_l$ and the distribution function $f_{l,m} = f(\mathbf{r}_l, \mathbf{k}_m, \tau)$, which is estimated with the help of (18) by introducing a mesh $\phi_m$ in the wavevector space, $(\Omega_{l,m} = \psi_l \phi_m)$, as

$$f_{l,m} = \frac{\sum_n \theta_{\Omega_{l,m}}(n)}{V_{\Omega_l} V_{\phi_m}} \quad C_l = \sum_m f_{l,m} V_{\phi_m} N_\tau = \sum_l C_l V_{\psi_l} \qquad (19)$$

The charge density $C_l$ is used to find the solution of the Poisson equation, which provides an update for the electric force $\mathbf{F}(\mathbf{r},t)$. The latter governs the trajectories evolving the particles in the next time interval $(\tau, \tau + \Delta t)$. Between the steps of solving the Poisson equation the electric field is *frozen* so that event biasing can be applied. Asssume that at time $\tau$ the particles emerge with weights $W_n$. Due to the event biasing the behavior of the biased particles differs from that of the EMC particles. The distribution function of the biased particles $f_{l.m}^{num}$ obtained from the above formula is entirely different from $f_{l,m}$. Nevertheless, as seen from (15), any biasing does not change the values of the physical averages. The Boltzmann distribution function is recovered by using the weights $W_n$ as

$$f_{l,m} = \frac{\sum_n W_n \theta_{\Omega_{l,m}}(n)}{V_{\Omega_l} V_{\phi_m}}. \qquad (20)$$

Accordingly, the correct **F** is provided by the Poisson equation. As the evolution is Markovian, $f_{l,m}$ presents the initial condition for the next time step. Numerical particles, having distribution $f_{l,m}^{num}$ and weights $W_n$ present a biased initial condition for this step. The initial weight will be updated in the time interval $(\tau, \tau+\Delta t)$ by the weight factors according to the chosen biased evolution. It follows that the particle weights *survive* between the successive iteration steps, which completes the proof of the self-consistent biasing scheme.

## 4.2 Biasing Methods

Three different event-biasing techniques have been employed in this work. The chosen biasing techniques aim at increasing the number of *numerical* particles in the channel, but keep the total particle number in the device constant (for example, equal to $10^5$), as used in the EMC technique. Particles, which enrich the high energy domain of the distribution on the expense of obtaining weights less than unity readily overcome the source potential barrier. Particle number in the low energy domain is less than the conventional ones, with higher weights and remain longer in the S/D regions. The methods are discussed in the following:

**(a)** *Biasing the Initial/Boundary Temperature*: Denoting the equilibrium distribution,

$$f_{eq}(\varepsilon, T) = \frac{1}{\overline{\varepsilon}} e^{-\frac{\varepsilon}{\overline{\varepsilon}}}, \tag{21}$$

where, $\varepsilon$ is the individual carrier energy and $\overline{\varepsilon} = 1.5 k_B T$, one chooses a biased distribution

$$f_0^b(\varepsilon, T_b) = \frac{1}{\overline{\varepsilon}_b} e^{-\frac{\varepsilon}{\overline{\varepsilon}_b}}, \tag{22}$$

which corresponds to higher temperature $T_b = bias \times T$ (*bias* > 1). Having higher kinetic energy, the numerical particles readily overcome the source potential barrier and enrich the statistics in the channel. The weight distribution is governed by the formula

$$weight = bias \cdot \frac{\exp(-\varepsilon/\overline{\varepsilon})}{\exp(-\varepsilon/\overline{\varepsilon}_{bias})}. \tag{23}$$

The biasing scheme is illustrated in Figure 1. Increasing $T_b$ increases the spread of the weight further away from unity which may lead to an increased variance of the physical averages, obtained by the mean of heavy and light particles. Thus finding the appropriate bias is a matter of compromise between the need for more particles in the channel (high temperature) and keeping the spread of the weight low (low temperature). The particle weight distribution for a particular choice of *bias* (= 1.5 corresponding to a temperature of 450K) is shown in Figure 2(a). That biasing increases the numerical particle number in the channel region by decreasing the same in the S/D regions in illustrated in Figure 2(b) with *bias* = 3. Also noticeable is the fact that different values of *bias* does not change the actual number of real (physical) particles throughout the device region.

**(b) *Particle Split*:** Particle weight is controlled by choosing a desired weight $w1$ of the numerical particles with kinetic energy below given level $\varepsilon 1$ and weight $w2$ with kinetic energy above $\varepsilon 1$. $f^b$ is obtained from $f_{eq}$ as follows:

$$f_0^b(\varepsilon) = \frac{f_{eq}(\varepsilon)}{w1}, \quad \varepsilon \leq \varepsilon_1$$
$$f_0^b(\varepsilon) = \frac{f_{eq}(\varepsilon)}{w2}, \quad \varepsilon > \varepsilon_1 \quad .$$
(24)

$w2$ is obtained as a function of $w1$ and $\varepsilon 1$ from the condition for normalization of $f^b$:

$$w2 = \frac{w1 \cdot e^{-\frac{\varepsilon}{\bar{\varepsilon}}}}{w1 - 1 + e^{-\frac{\varepsilon}{\bar{\varepsilon}}}}.$$
(25)

A choice of $w1 > 1$ effectively reduces the number of particles below $\varepsilon 1$ as compared to the unbiased case. In both the above cases of biasing heavy particles which enter the channel perturb the statistics accumulated by a set of light weight particles (Figure 3). It is thus desirable to apply the technique of particle splitting in parallel to the temperature biasing in order to minimize the spread of the weight.

**(c) *Biasing Phonon Scattering (e-a)*:** Artificial carrier heating can be achieved by biasing the phonon scattering rates. For a given scattering mechanism, the probability for phonon absorption is increased at the expense of phonon emission, controlled by a parameter $w1$,

$$\lambda_{em}^b = \frac{\lambda_{em}}{w}, \quad w > 1$$
$$\lambda_{abs}^b = \lambda_{abs} + \left(\lambda_{em} - \frac{\lambda_{em}}{w}\right) \quad .$$
(26)

If in the course of the simulation, a phonon absorption is selected, the particle weight is updated by a multiplication with $\lambda_{abs}/\lambda_{abs}^b$, otherwise with $\lambda_{em}/\lambda_{em}^b$. The distribution of the flight time is not affected, because the sum of emission and absorption rate is not changed.

### 4.3  Simulation results

The MOSFET device, chosen for the simulation experiments has gate length of 15 nm, channel doping $2\times 10^{19}$ cm$^{-3}$, and oxide thickness 0.8 nm. Similar device has already been fabricated by Intel [27] . The applied potenitals $V_G$ = 0.375V, $V_D$ = 0.1V correspond to a subthreshold regime at lattice temperature $T$ = 300K.

Figure 4 shows the standard deviation in electron number as a function of evolution time in a certain cell within the channel region. Clearly one can see the improvement from the use of event biasing techniques. The standard deviation is least for method (a) with the initial/boundary temperature being biased.

Regarding the validation, *first*, the consistency of the biasing techniques in the thermodynamic limit of a very large number ($10^5$) of simulated particles is investigated. Both Boltzmann and biased stochastic processes must give the same evolution of the physical averages. Figure 5 shows that the biased experiments recover precisely the physical averages of sheet electron density and electron energy along the channel of the device. *Second*, investigated is the convergence of the cumulative averages for the *channel* and *terminal* currents obtained from the velocity and particle counting, respectively. Biasing the phonon scattering rates (e-a) is applied in the half of the source region near the barrier in a 4 nm depth. Figure 6 (top panel) shows the biased *channel current* as compared to the EMC result for 30 ps evolution time. The 5% error region (straight lines) around the mean value is entered 2.5 ps earlier and the convergence is better. The *channel current* from the boundary temperature biasing ($T$ = 450K) is shown in the bottom panel of Figure 6. The temperature-biased curve shows a superior behavior. The corresponding *terminal currents* (shown in Figure 7) are much more noisy and show long-time correlations due to the inter-particle interactions. The e-a bised curve is very unstable and enters the 5% error region in Figure 7 (top panel), after 15 ps evolution. One can associate this behavior with the numerical error. The poor statistics is due to the appearance of very heavy $(W>2)$ particles in the source as can be seen in Figure 8(a). To check this, it is sufficient to apply the conventional particle *splitting* coupled with the e-a biasing. The result is presented by the dotted curve in Figure 7. The behavior is significantly improved on the expense of a 30% increase of the simulated particles (the corresponding weight distribution is shown in Figure 8(b)). The terminal current corresponding to the biasing of the temperature of the injected particles again shows a superior behavior. This is due to an improved weight control. The weight, determined during the injection remains constant in the evolution. Its maximal value for $T = 450K$ is exactly 1.5 as depicted in Figure 2(a). Furthermore the probability for interaction with the impurities, which dominates the S/D regions, drops for the majority of the particles due to their high energies. The conventional splitting technique cannot achieve such superiority. The terminal current from *particle split* technique is shown by the dotted curve on Figure 7 (bottom panel). The behavior of the curve resembles the EMC counterpart. An improvement is expected if the *w*2 particles are additionally split, which recovers the *conventional* split technique.

## 5. Conclusion

In conclusion, the event biasing approach has been derived in presence of both initial and boundary conditions and generalized for self-consistent simulations. The approach is confirmed by the presented simulations. A bias technique, particularly useful for small devices, is obtained by injection of hot carriers from the boundaries. The coupling with the Poisson equation requires a precise statistics in the S/D regions. It is shown that a combination of event biasing and population control approaches is advantageous for this purpose.

**Acknowledgements**

The authors would like to thank Dragica Vasileska for useful discussions.

# LIST OF FIGURES

*Islam et al.*

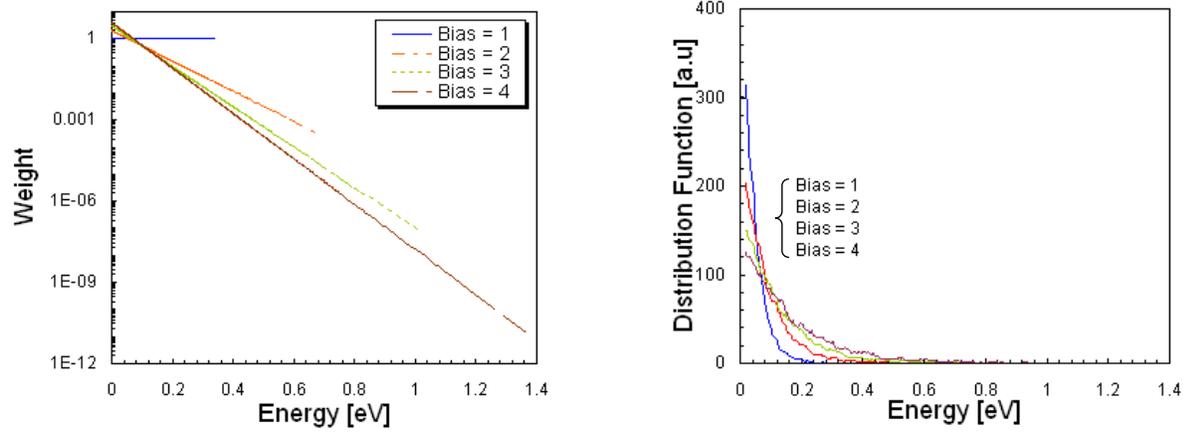

Figure 1. A weighting scheme used in the event-biasing method.

*Islam et al.*

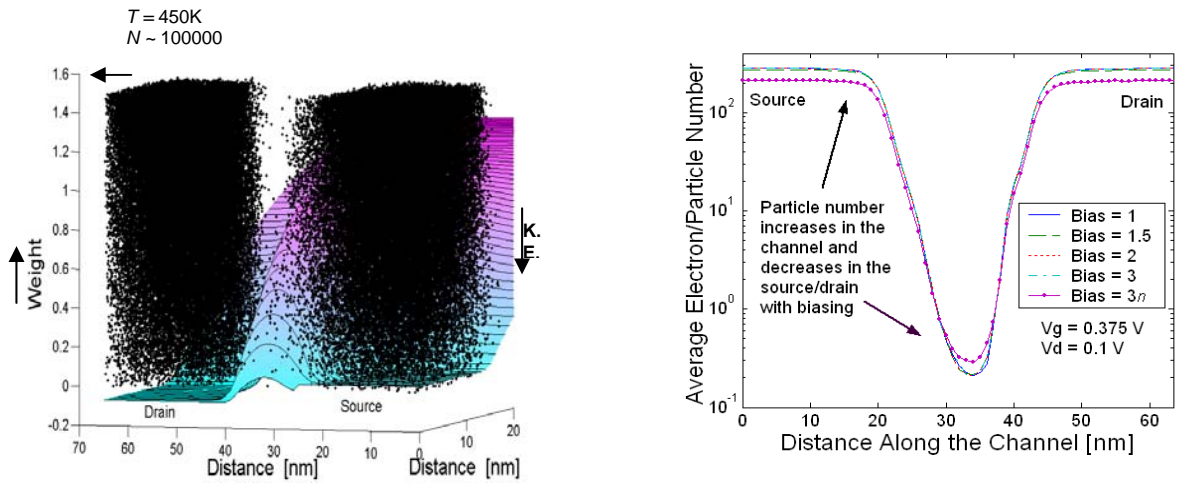

Figure 2. (a) Weight distribution for $T$ = 450K, and (b) Biasing decreases numerical particle number in the source/drain regions while increases the number in the channel region.

*Islam et al.*

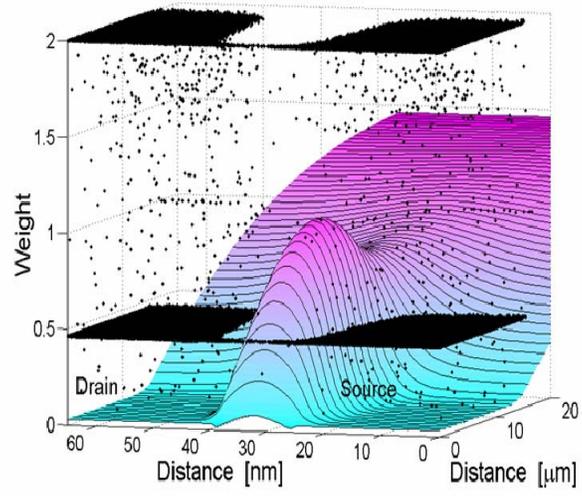

Figure 3. Particle split method and the distribution of particles.

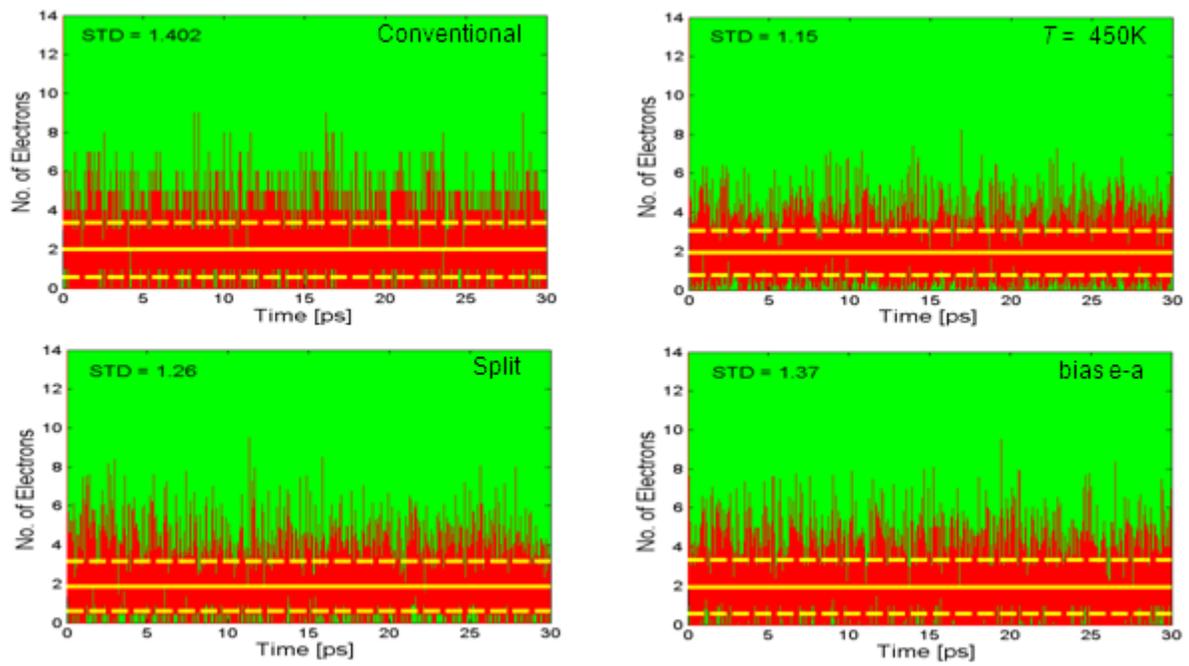

Figure 4. Enhancement of channel statistics: reduction of standard deviation.

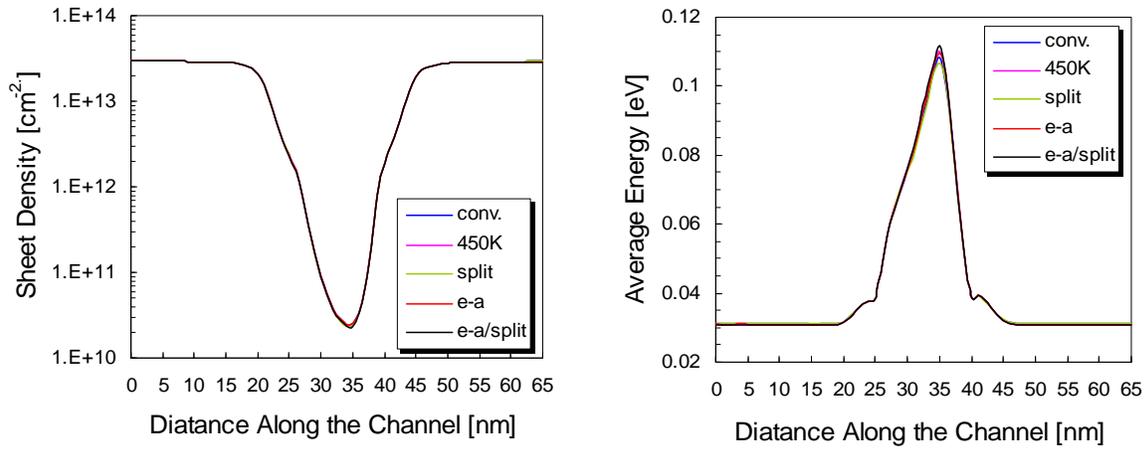

Figure 5. Biasing recovers precisely the self-consistent average sheet density (left panel) and average kinetic energy of the electrons (right panel).

*Islam et al.*

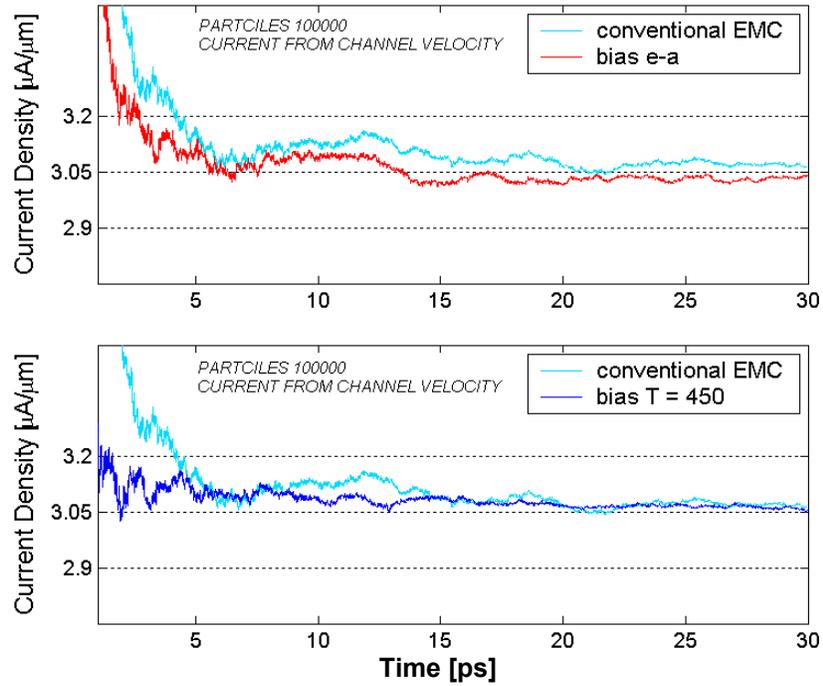

Figure 6. Comparison of the *channel currents* obtained from (1) biased e-a rates (top panel), and (2) biased boundary distribution (bottom panel) methods from velocity consideration.

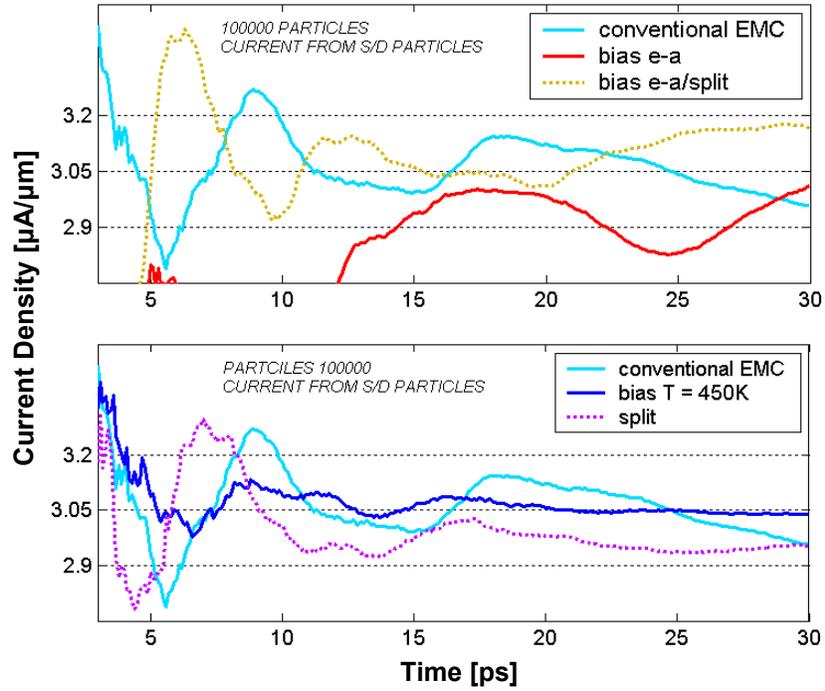

Figure 7. *Terminal currents* obtained from various methods by particle counting.

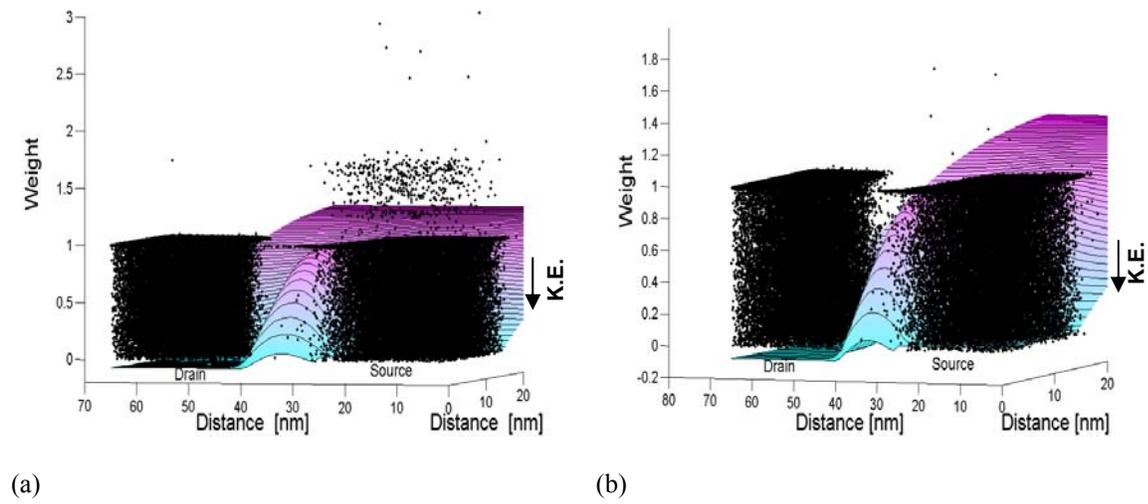

(a)　　　　　　　　　　　　　　　　　(b)

Figure 8. Numerical particle weight distribution in (a) e-a biasing, and (b) e-a/split biasing.